\documentclass[prl,twocolumn,showpacs]{revtex4-1}

\usepackage{graphicx}
\usepackage{ulem}
\usepackage{dcolumn}
\usepackage{bm}
\usepackage{natbib}

\begin{document}

\title{New Interpretation for the Observed Cosmological Redshifts and its Implications}

\author{Branislav Vlahovic}
\email{vlahovic@nccu.edu}
\affiliation{Department  of Physics, North Carolina Central University, 1801 Fayetteville 	Street, Durham, NC 27707 USA.}


\begin{abstract}
\noindent
The cosmological redshifts $z$ in the frequencies of spectral lines from distant galaxies as compared with their values observed in terrestrial laboratories, which are due to the scale factor $a(t)$, frequently are interpret as a special-relativistic Doppler shift alone. We will demonstrate that this interpretation is not correct and that the contribution of the gravitational redshift is always present and significant.
We will show that the gravitational redshift is actually about the same magnitude as the cosmological redshift, but that only for cosmological models without the dark energy component cosmological and gravitational redshift can be considered to be the same. Significant contribution of the gravitational redshift due to the gravitational field of the Universe, which is ignored in interpretation of the observational data, could have significant impact on cosmological theories. We will first calculate contributions of gravitational redshift to CMB and time dilation of Type Ia supernovae, and use it to explain the excess redshifts of quasars and active galaxies, and redshifts of companion galaxies of stars.  We will show its possible implications on the interpretation of mass density of matter and mass as function of cosmological time. Finally we will demonstrate that taking into account gravitational redshift allows to interpret luminosity distance and surface brightness of distant galaxies to be consistent with the static universe cosmological models.
\end{abstract}
\maketitle

\section{Background}

 In co-moving Robertson-Walker coordinate system relation between the redshift $z$, in the frequencies of spectral lines from distant galaxies as compared with their values observed in terrestrial laboratories, and scale factor $a(t)$ is given by
\begin{equation} \label{redshift}
  1+z = a(t_0)/a(t_1),
\end{equation}
where the light is emitted at time $t_1$ and observed at the present time $t_0$. Such redshift $z$ is frequently interpreted in terms of the Doppler effect. The reason for that is that for a decreasing or increasing $a(t)$, the proper distance to any co-moving light source decreases or increases with time, so that such sources are approaching us or receding from us. For this reason the galaxies with wavelength shift $z$ are often said to have a cosmological radial velocity $cz$.

	However there is a problem with interpretation of the cosmological redshift as a Doppler shift. First, the change in wavelength from emission to detection of light does not depend on the rate of change of $a(t)$ at the times of emission or detection, but on the increase of $a(t)$ in the whole period from emission to absorption. Therefore we cannot say anything about the velocity of a galaxy at time $t_1$ by measuring redshift $z$, nor we should express the radial galaxy velocity through $cz$. The observed redshift, wavelength stretching, is caused by stretching of the space not by the velocity of the galaxy at time $t_1$. The galaxy could have any velocity at that time and we cannot measure it, because we observe just the cumulative effect in wavelength change during the time period $t_1-t_0$. However, in series of papers \cite{1R}, in the 1920's, Wirtz and K. Lundmark showed that Slipher's redshifts, summarized in \cite{Slipher}, increased with the distance and therefore could most easily be understood in terms of a general recession of distant galaxies.

	Another problem is that the wavelength of light is also affected by the gravitational field of the universe, gravitational redshift. For that reason it is not correct to interpret the redshift from very distant sources in terms of cosmological redshift alone or to interpret it as a special-relativistic Doppler shift alone, the gravitational redshift must be taken into account too. One can claim that cosmological and gravitational redshift are the same and must be the same, that they are just different interpretation of the same effect. For instance, that it follows from the solution of the Friedmann equations.
This equivalence will be desirable because at the present there is a problem to explain the physical mechanism that causes the cosmological redshift. The understanding  is that space is expanding and that it stretches the wavelength. However, the energy of the photons is changed and to change the energy there should be an interaction, a force between space which is expanding and the photons, which will be responsible for that change in photons energy. There is no a known mechanism that will describe such interaction. On the other hand, there is no such problem if the cosmological redshift can be interpreted as gravitational and using quantum mechanical interpretation of photon.
However, as it will be shown later the equivalence between cosmological and gravitational redshift is only true for cosmological models without dark energy. Since all current models assume dark energy as a dominant component all redshifts data must be reinterpreted.

The gravitational redshift has been always associated only with strong local gravitational fields. For that reason it has been ignored, because for instance for a typical cluster mass of  $\sim 10^{14}M_\odot$, where $M_\odot$ is the Sun's mass, the gravitational redshift is estimated \cite{2R}, \cite{3R}, \cite{4R} to be $\approx$ 10 km$s^{-1}$, which is about two orders of magnitude smaller than Doppler shift due to the random motion of galaxies in cluster.  However, the gravitational redshift is experimentally verified first in \cite{2a} and more recently in \cite{2b} and also it is recently  measured in galaxy clusters \cite{5R}. We will show that gravitational redshift due to the gravitational field of the universe, which is different from that of a local gravitational field, is not small and cannot be ignored. It is actually of the order of the cosmological redshift.

We will show that it is necessary to take into account gravitational redshift when light is emitted at an early epoch and observed at the present time, because of the change in gravitational potential, due to the expansion of space and because light is losing energy by climbing trough the gravitational field. We will apply gravitational redshift to solve some current problems in interpreting observable data.

\section{Gravitational Redshift }


One of the central predictions of metric theories of gravity, such as general relativity, is that photons will lose energy leaving a massive object and gain it when moving toward a gravitational source. The redshift between two identical frequency standards placed at rest at different strengths in a static gravitational field is:
\begin{equation} \label{redshift__1}
  \Delta\nu/\nu = -\Delta\lambda/\lambda = \frac {\Delta U} {c^2} = \frac {GM_1} {R_1c^2} - \frac {GM_0} {R_0c^2},
  \end{equation}
where $R_1$ is for instance radial position of galaxy that emmited the photon at time $t_1$ and $R_0$ is the radial position of observer that detected the emitted photon at present time $t_0$. If we assume that the Universe is a sphere then $M_1$ and $M_0$ are the portions of the mass of the Universe inside the imaginary spheres going through coordinates $R_1$ and $R_0$ respectively. For instance if we would like to calculate gravitational redshift for CMB then $R_1=42$ Mly, which is the size of the Universe at the time of decoupling \cite{2bb}, $R_0=46$ Gly  is the size of visible Universe at the present time  \cite{WMAP}, and $M_1$=$M_0$=$M$ is the mass of the Universe, which is the same at time of decoupling and today.

 Using  for the mass of the visible Universe $M_1$=$M_0$=$2\times10^{53}$ kg, we obtain for the CMB redshift due to the gravitation $z$=371. However,
 if we use a plausible model of the Universe described in \cite{Vla}, we will in that case obtain for $z$ values around 1166, which is in agreement with observed value, taking into account uncertainty in the $M$.

\begin{figure}[h]
		\includegraphics[width=7.5cm]{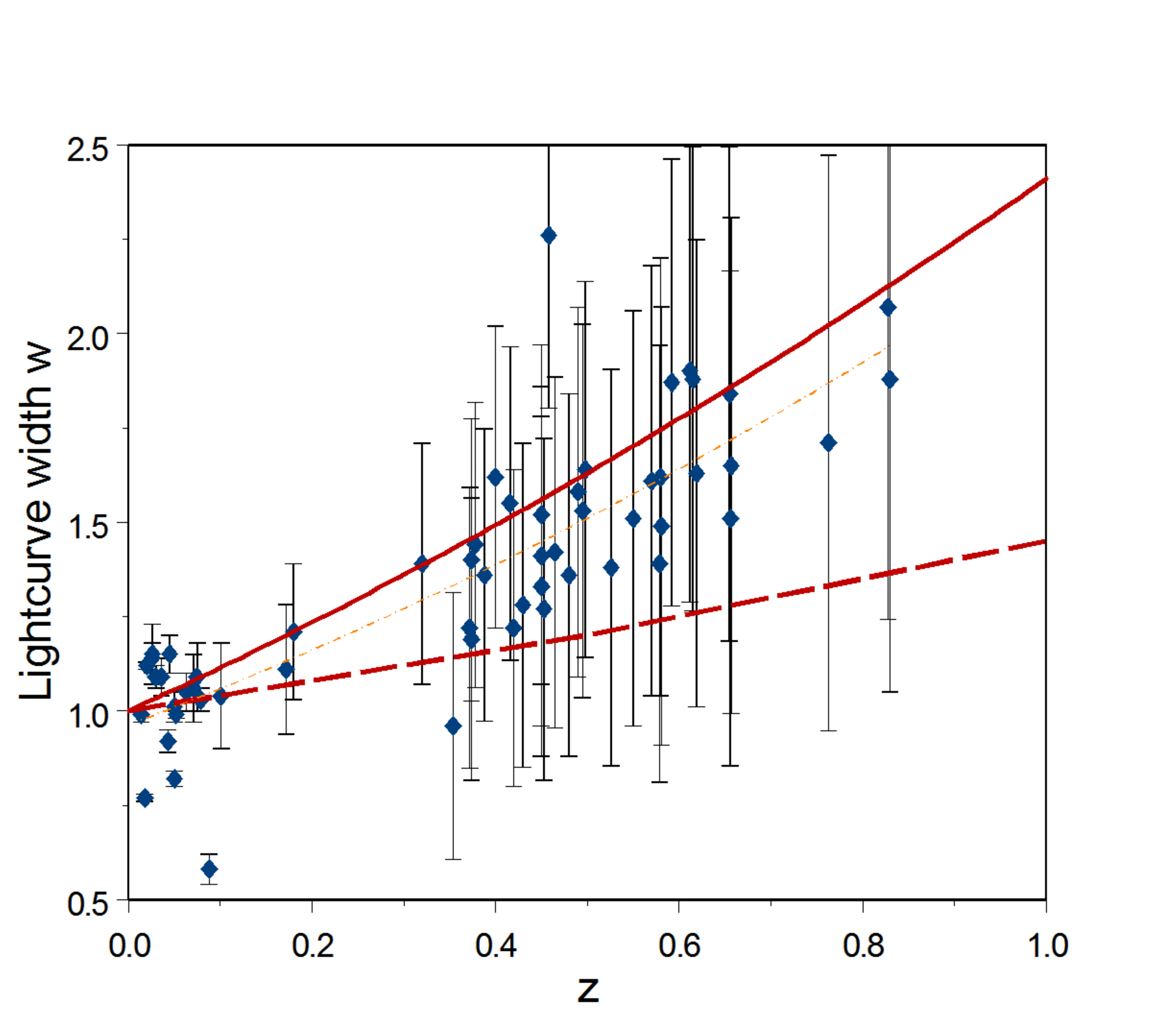}
		\caption{\label{fig-time}
The time dilatation of Ia supernovae. Dashed line are calculations for different z using equation (\ref{time}) and baryonic mass only. Solid line is for the cosmological model \cite{Vla}.  The experimental data \cite{2c} are shown by rhombus. The fine dot-dashed line is our least square fit of the data.
}
\end{figure}

Let us now consider the effect of the gravitation field of the Universe on time dilation of Type Ia supernovae (SNe Ia), because a clock in a gravitational potential U will run more slowly by a factor $\Delta t$
\begin{equation} \label{time}
  \Delta t  = 1+\Delta U/c^2,
  \end{equation}
 as compared to a similar clock outside the potential. Calculating for $z$=0.5 gives a factor of 1.20 and for $z$=1 a factor of 1.45, if we take into account only baryonic mass $M$=$2\times10^{53}$ kg, and values 1.63 and 2.41 for a cosmological model that assumes a smaller  size of Universe for factor $\pi$ \cite{Vla}.
  In both cases the results are inside of one standard deviation with results \cite{2c}, see Fig. \ref{fig-time}.
It is important to note that in our calculations, for the CMB redshift and for time dilation, we used in (\ref{redshift__1}) and (\ref{time}) for $M$ the total mass of the Universe, as we are on the top of the shell representing the Universe, while we may be somewhere inside the shell. So, there is some allowed range for the obtained results, which depends on our position relative to the mass distribution in the Universe. Obviously, even if we take for the mass of the Universe only baryonic mass, in both cases of CMB and Type Ia supernovae, the contribution of the gravitational redshift due to the gravitational field of the Universe is significant and should be taken into account in data interpretation, for instance in interpreting the acceleration of Universe expansion using supernovae data or redshifts of QSOs.

 An argument that we do not need to take into account gravitational redshift in interpreting observational data could be that it is equivalent to the cosmological redshift. One can say that we did not obtained accidentally the same values for CMB and Type I Sn supernova, using equations (\ref{redshift__1}) and (\ref{time}), as it is predicted with cosmological redshift, that actually we must obtain exactly the same values if calculations are correctly done. That can be easily checked by evaluating the Friedmann energy equation
\begin{equation} \label{FE}
  (\frac {1} {a} \frac {da} {dt})^2 =  \frac {8\pi G}{3}\rho -  \frac {kc^2} {a^2}.
  \end{equation}
Comparing results obtained by solving the Friedmann equation for $k$=0 (for simplicity) and equation (\ref{redshift__1}) it is straightforward to show that in the case of baryonic mass the cosmological and the gravitational redshifts are the same. However, it is also obvious that in the model that includes only dark energy, the Friedmann equations will give cosmological redshift, while equation (\ref{redshift__1}) for the gravitational redshift will give actually a gravitational blue shift (because of the repulsion).  In standard $\Lambda$CDM model dark energy is dominant and therefore cosmological and gravitational redshift will be different, which should be taken into account in data interpretation. It is also possible to use (\ref{redshift__1}) to obtain a fine tune of the $M/R$ ratio for different cosmological epochs, since calculated gravitational redshift, using equation (\ref{redshift__1}), should agree with observed redshift.

As it is shown above taking into account only baryonic mass gives the redshift about 371 and 1166 for the current model of the visible Universe \cite {WMAP} and model proposed in \cite{Vla} respectively, which is close to the observed redshift. However, if we take in addition into account also dark mass, equation (\ref{redshift__1}) will give redshifts that are for factor 7 higher, which is not acceptable result, significantly different from observed.

The time dilation (\ref {time}) can be obtained for instance by comparing proper time intervals $\tau$ for clocks 1 and 2 placed at finite distances $R_1$ and $R_2$
\begin{equation} \label{Clocks}
\frac {d\tau_2} { d\tau_2} = \frac {\sqrt{g_{00}(2)dt}} {\sqrt{g_{00}(1)dt}} \simeq 1 + \Phi_2 - \Phi_1 = 1 - \frac {GM} {R_2c^2} + \frac {GM} {R_1c^2}.
\end{equation}
One can say that gravitational time dilation has purely geometric interpretation due to the metric tensor $g_{00}$, that modification of the clock rate is due to the change in spacetime geometry. However, it does not mean that the time dilation caused by cosmological expansion, obtained by solving (\ref{FE}), will be the same as the result of equation (\ref{Clocks}).  Gravitational redshift assumes that the material content of the Universe determines the preferred inertial motion, that the Hubble flow, recession of the enormous amount of the matter causes the redshift and time dilation. That is not so in general relativity because Birkhoff's theorem is the basis for the equation
  \begin{equation} \label{g}
   \nabla \cdot {\bf{g}} = 4\pi G(\rho + 3p)
 \end{equation}
where  density $\rho$  and pressure  $p$ represent active gravitational source for gravitational acceleration. Assumption is that into flat spacetime we placed enough small amount of active gravitational mass $\rho+3p$ so that we can use Newtonian mechanic and equation  (\ref{g}). Also when we are solving equation (\ref{FE}) we are expressing $p$ by $\rho$ assuming that the net energy within the sphere of volume $V$ is $U=\rho V$, which neglects the gravitational energy of the mass within the sphere. However, neither of these two assumptions is correct when the considered volume is the entire Universe and the active gravitational mass is the mass of the Universe.  For that reason the redshifts and time dilation due to the cosmological expansion calculated using equation (\ref{FE}) and gravitational redshifts and time dilation obtained by equations (\ref{redshift__1}) and (\ref{Clocks}) will be different.

The gravitational redshift can be explained by the energy conservation and by the principle of equivalence. Consider for example an atom in gravitational field $\Phi$ at rest, emitting photon $h\nu_1$  and the same photon absorbed by another atom at rest, at a different position. The mass lost by the first atom by emission of the photon is $\Delta m_1 = -h\nu_1$ and the mass gained by the another atom by absorption of the photon is  $\Delta m_2 = h\nu_2$. In emission absorption process the total energy must be conserved:
\begin{equation} \label{COE}
\Delta m_1 + \Phi_1 \Delta m_1 + \Delta m_2 + \Phi_2 \Delta m_2  = 0
\end{equation}
where we used that the energy change $\Delta m$ produces the change in the gravitational mass, due to the equivalence principle.

This gives for the gravitational redshift
 \begin{equation} \label{RG}
  \Delta \nu /\nu =  \frac {\nu_2 - \nu_1} {\nu_1} \simeq  \Phi_1 - \Phi_2
  \end{equation}
The equation (\ref{COE}) is conservation of energy that must be satisfied for any model and it must be satisfied not only for photons and radiation mass, it also must be satisfied for baryonic mass and dark mass. This has significant implications. It means that not only particles of radiation will be slow down during expansion of the Universe, but also that particles of matter will also lose energy. As it can be seen from (\ref{COE}) the matter mass and therefore the mass density of mater will be affected on the same way as the radiation density. It means that the density of matter $\rho_m$ and density of radiation $\rho_r$ must have the same expansion dependence a(t):
\begin{equation} \label{DMR}
\rho_r \propto 1/a^4  \ \ \ \ and \ also  \ \ \ \ \rho_m \propto 1/a^4.
  \end{equation}
This will have significant impact on cosmological models, because the interpretation of observed data will be different if  $\rho_r \propto 1/a^4$ instead to $1/a^3$.

An additional interesting consequence of relation (\ref{COE}) is that in an expanding universe the mass will depend on the epoch, because it will change accordingly to change in the gravitational potential.
For the same reason that light is losing energy climbing through the gravitational field, on the same way the matter will also lose energy. This can be expressed through the mass energy relation by a lower mass.  It follows from (\ref{COE}) that the expansion of the space requires space-time dependent mass.
\begin{equation} \label{MD}
\Delta m(a) \propto \frac {1} {a_1(t)} -  \frac {1} {a_2(t)}
  \end{equation}
The matter participating in Hubble flow will change mass by the flow, because it slows down by climbing in the gravitational potential, which is changing by expansion.
The mass of the the matter at the early epoch (for instance the time of decoupling) will be larger than the same amount of mass at the present time, because of the stronger gravitational potential at the earlier epoch than at the present. It is important to note that this is different from the variable mass hypothesis proposed in 1977 by Narlikar \cite{Narlikar6R}, where the mass is proportional to the time of creation $m\propto t^2$. In our case, if there is no Hubble flow, for instance in case of a static universe, the mass will be constant in time. Also the proportionality is different, in \cite{Narlikar6R} mass is smaller in early epoch, while gravitational redshift requires that due to the energy lose during the expansion it should be smaller as Universe expands.

One can argue that this will cause an additional redshift, actually a blueshift, because the light emitted by the matter that has bigger mass will have shorter wavelength in comparison by the light emitted by a smaller mass. This is the same mechanism as proposed by \cite{Narliker-Arp}. However at the presence of bigger mass time is dilated, which will cause the redshift of the same magnitude. These two effects cancel each other; the only gravitational effect that will remain is due to particle climbing up in the gravitational potential.

There are observable indications that quasars of generally larger redshifts are associated with larger galaxies of much lower redshift \cite{Arp 1990}.  That and the apparent lack of the time dilation effect in quasars light curves was used as argument that quasars are not at the cosmological distances implied by their redshifts \cite{Arp Russell}. The most recent results \cite{Kazuhiro} on the origin of the radio jet in M87 at the location of the central black hole, implies that the site of material infall onto the black hole and accretion disk and the eventual origin of the jets is just about 0.007 to 0.01 pc or 14-23 Schwarzschild radii ($R_s$ radius of the event horizon). This is much closer than previous thought, $10^4$-$10^5$ $R_s$. The reveal that the central engine of M87 is located in the scale of 10 $R_s$ of the radio core is in synergy with observation with very-high-energy (VHE) $\gamma$ rays. The recent observations of an intense VHE $\gamma$ -rays flare of M87 suggest that the VHE emission originates in the core and that the size of the VHE emission region is also on the scale of 10 $R_s$ \cite{Acciari}.  Thus the new results indicate much higher gravitational potential, ratio $M/R$, that emitted radiation needs to overcome by climbing up in the gravitation field to escape from the quasar. This allows to interpret the observed redshift not as cosmological only, but with significant gravitational redshift component, which in some cases may even be larger than cosmological redshift.  For example, in the case of M87 using for $M$=$6\times10^9$ $M_\odot$ \cite{Gebhardt} and $R$=23$R_s$ (0.01 pc) \cite{Kazuhiro} the gravitational redshift is of the order of $z$=0.05. It is even bigger than the observed value of $\sim$ 0.005 (16.7 Mpc), which demonstrates a significant contribution of the gravitational redshift. Hoverer, the obtained result is in agreement with the measured redshift because there is still freedom in the estimate of $M/R$ ratio. Therefore, it is quite possible that discrepancies between galaxies with a small cosmological redshift (the gravitational redshift is not significant because of small $M/R$, large size of galaxy) and an accompanied quasar with a large redshift (large $M/R$) can be explained by gravitational redshift.  It is true that this will not solve the problem of the time dilation (because it should be also present in gravitational redshift), but it solves problem with the difference in the redshifts associated with galaxy and a nearby quasar.

There is also established excess redshift of companion galaxies and stars, grouped together around a large dominant galaxy. In two nearest groups, the local group and M81 group, the companion galaxies have redshift systematically larger by order of 100 km $s^{-1}$. The effect is statistically significant \cite{Arp 1990}, \cite{Arp 1987}, but until now not explained. However, the explanation can be done by gravitational redshift. It is enough to assume that in the past companion galaxies were closer to the dominant galaxy and that for some reason, for instance initial motion, or because of the excess of rotation motion that overcome the gravity force, they are at present at larger distances. By moving up in the gravitational field of the central galaxy all companion galaxies will be redshifted. Because the observed redshifts are small only about 100 km $s^{-1}$ they can be explained by (\ref{redshift__1}) and galaxy $R/M$.

It is interesting to note that gravitational redshift also gives possibility for the static universe as a reasonable option.
 Actually, the amount of gravitational contribution to the redshift, in both cases CMB and supernovae, is so significant that it allows to interpret the data solely as the gravitational effect, without cosmological component $a(t_0)/a(t_1)$ related to the expansion. This now raises question can we have a static cosmological model that will be consistent with the observational data. A model without cosmological redshift, but gravitational one.

 The main reason that discredits the static solution, for instance {\it tired light model} is that it cannot describe correctly relation between luminosity distance $d_L(z)$ and diameter distance $d_A(z)$ of the same distant source. The model independent interpretation of redshift and luminosity require $D_L(z)/d_A(z) = (1+z)^{-2}$, while static models predict factor $(1+z)^{-1}$. In conventional theory all rates of the sources are described by a factor $(1+z)^{-1}$, while in tired light model there is no such slowing down. However, this statement is not correct. It is true that there is no cosmological redshift $a(0)/a(t)$, because there is no expansion. However, there is gravitational redshift. If we imagine Universe as a uniform homogenous static sphere, light will be gravitationally redshifted, in the case of radial motion of light from smaller toward larger radii. Please note that not only the wavelength of the photon will be affected, but also because of time dilation (due to gravitation) the rate of the photon will be also reduced.  So the static model cannot be disqualified.

One can say that in this case one can also expect blueshift, if we are not on the surface of the sphere representing the Universe. It is actually possible that we observe these blue shifted photons. The objects like $S_\mathit{b}$ galaxy M31 have a slightly negative redshift (-86 km $s^{-1}$), which cannot be velocity without violating the dynamics of our Local Group \cite{Arp 1986}. There are also $S_\mathit{b}$ galaxies in Virgo cluster that are blueshifted. So it is possible that those galaxies are in outer shell, and that it is the reason for gravitational blueshift.

\section{Conclusion}

The new interpretation of the observed redshifts shows that gravitational redshift and cosmological redshift are the same only for some specific cosmological models, which do not include dark energy. The gravitational effects due to the Universe expansion and large gravitational potential could be significant and in some cases can be larger than cosmological redshift and for that reason should be taken into account in data analysis.  The gravitational redshift affects radiation matter and particles of matter, which require the same $1/R^4$ dependence for both, radiation and particle matter. It also gives possibility that mass depends on space expansion and epoch. The gravitational redshift gives reasonable explanation for excess redshift of quasars and active galaxies. It also explains excess redshift of companion galaxies and stars. Finally it allows for a static model of the Universe, because for that model it predicts time dilation, which lack of prediction by the cosmological interpretation of the time dilation was the main argument against the static model.

\begin{acknowledgments}
I would like to thank S. Matinyan and I. Filikhin for useful discussions. This work is supported by NSF award HRD-0833184 and NASA grant NNX09AV07A.
\end{acknowledgments}

\end{document}